\documentclass[10pt,conference,compsocconf]{IEEEtran}
\usepackage{times}

\usepackage{caption}
\ifCLASSINFOpdf
   \usepackage[pdftex]{graphicx}
\else
\fi
\begin{document}
\pagenumbering{gobble}
%
\title{Trie Compression for GPU Accelerated Multi-Pattern Matching}

\author{\IEEEauthorblockN{Xavier Bellekens}
\IEEEauthorblockA{Division of Computing and Mathematics\\
Abertay University\\Dundee, Scotland\\
Email: x.bellekens@abertay.ac.uk}
\and
\IEEEauthorblockN{Amar Seeam}
\IEEEauthorblockA{Department of Computer Science\\
Middlesex University\\
Mauritius Campus\\ 
Email: a.seeam@mdx.ac.uk}
\and
\IEEEauthorblockN{Christos Tachtatzis \& Robert Atkinson}
\IEEEauthorblockA{EEE Department\\
University of Strathclyde\\
Glasgow, Scotland\\
Email: name.surname@strath.ac.uk}
}

%


\maketitle

\begin{abstract}
Graphics Processing Units (GPU) allow for running massively parallel applications offloading the Central Processing Unit (CPU) from computationally intensive resources. However GPUs have a limited amount of memory. In this paper, a trie compression  algorithm for massively parallel pattern matching is presented demonstrating 85\% less space requirements than the original highly efficient parallel failure-less Aho-Corasick, whilst demonstrating over 22~Gbps throughput. The algorithm presented takes advantage of compressed row storage matrices as well as shared and texture memory on the GPU.
\end{abstract}


\begin{IEEEkeywords}
Pattern Matching Algorithm; Trie Compression; Searching; Data Compression; GPU
\end{IEEEkeywords}

%
\IEEEpeerreviewmaketitle
\section{Introduction}
Pattern matching algorithms are used in a plethora of fields, ranging from bio-medical applications to cyber-security, the internet of things (IoT), DNA sequencing and anti-virus systems. The ever growing volume of data to be analysed, often in real time,  demands high computational performance.  \newline

The massively parallel capabilities of Graphical Processor Units, have recently been exploited in numerous fields such as mathematics~\cite{nasridinov2014decision}, physics~\cite{nakasato2009oct}, life sciences~\cite{7576464},  computer science~\cite{aparicio2014scalable}, networking~\cite{retvari2013compressing}, and astronomy~\cite{bedorf2012bonsai} to increase the throughput of sequential algorithms and reduce the processing time.  \newline

With the increasing number of patterns to search for and the scarcity of memory on Graphics Processing Units data compression is important. The massively parallel capabilities allow for increasing the processing throughput and can benefit applications using string dictionaries~\cite{martinez2016practical}, or application requiring large trees~\cite{navarro2016faster}. \newline

The remainder of this paper is organised as follows: Section~II describes the GPU programming model, Section~III provides background on multi-pattern matching algorithms while, Section~IV discusses the failure-less Aho-Corasick algorithm used within this research. Section~V highlights the design and implementation of the trie compression algorithms, while Section~VI provides details on the environment. The results are highlighted in Section~VII and the paper finishes with the Conclusion in Section~VIII.
\section{Background}

\subsection{GPU Programming Model}
In this work, an Nvidia 1080 GPUs is used along with the Compute Unified Device Architecture (CUDA) programming model, allowing for a rich Software Development Kit (SDK). Using the CUDA SDK, researchers are able to communicate with GPUs using a variety of programming languages. The C language has been extended with primitives, libraries and compiler directives in order for software developers to be able to request and store data on GPUs for processing. \newline

GPUs are composed of numerous Streaming Multiprocessors~(SM) operating in a Single Instruction Multiple Thread~(SIMT) fashion. SMs are themselves composed of numerous CUDA cores, also known as Streaming Processors~(SP).	

\subsection{Multi-Pattern Matching}
Pattern matching is the art of searching for a pattern $P$ in a text $T$. Multi-pattern matching algorithms are used in a plethora of domains ranging from cyber-security to biology and engineering~\cite{tran2011bit}. \newline

The Aho-Corasick algorithm is one of the most widely used algorithms~\cite{Aho:1975:ESM:360825.360855}\cite{vakili2016memory}. The algorithm allows the matching of multiple patterns in a single pass over a text. This is achieved by using the failure links created during the construction phase of the algorithm.\newline

The Aho-Corasick, however, presents a major drawback when parallelised, as some patterns may be split over two different chunks of $T$. Each thread is required to overlap the next chunk of data by the length of the longest pattern $-1$. This drawback was described in~\cite{bellekens} and~\cite{tumeo2010efficient}.

\subsection{Parallel Failure-Less Aho-Corasick}
Lin~\textit{et~al.} presented an alternative method for multi-pattern matching on GPU in~\cite{5683320}. \\

\begin{figure*}[!t]
  \includegraphics[width=6.9in]{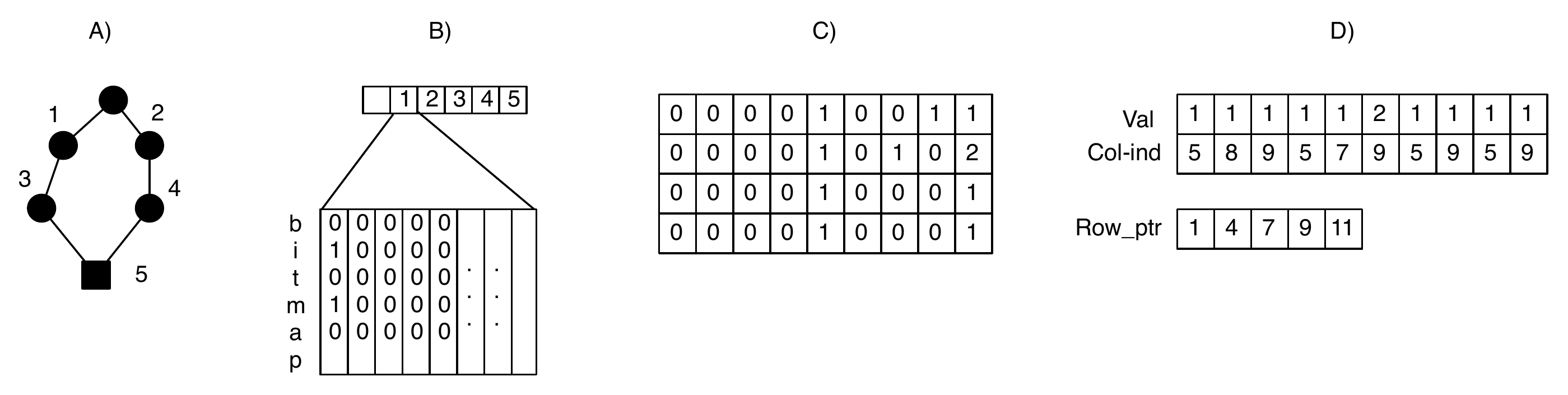}
  \caption{Memory layout transformation for a simplified transfer between the host and the device, allowing for better compression and improved throughput}
\label{fig:prefixMatching}
\end{figure*}
To overcome these problems, Lin~\textit{et~al.} presented the failure-less Aho-Corasick algorithm.  Each thread is assigned to a single letter in the text $T$. If a match is recorded, the thread continues the matching process until a mismatch. When a mismatch occurs the thread is terminated, releasing GPU resources. The algorithm also allows for coalesced memory access during the first memory transfer, and early thread termination.
\section{Design and Implementation}
The trie compression library presented within this section builds upon prior research presented in~\cite{Bellekens:2014:HMS:2659651.2659723}~and~\cite{bellekens2016high} and aims to further reduce the memory footprint of the highly-efficient parallel failure-less Aho-Corasick (HEPFAC) trie presented in~\cite{bellekens2016high}, while improving upon the tradeoff between memory compression operation and the throughput offered by the massively parallel capabilities of GPUs.  \newline

The compressed trie presented in our prior research is created in six distinct steps. I) The trie is constructed in a breadth-first approach, level by level. II) The trie is stored in a row major ordered array. III) The trie is truncated at the appropriate level, as described in~\cite{Bellekens:2014:HMS:2659651.2659723}. IV) Similar suffixes are merged together on the last three levels of the trie (This may vary based on the alphabet in use). V) The last nodes are merged together. VI) The row major ordering array is translated into a sparse matrix as described in~\cite{bellekens2016high}.\newline

\begin{figure}[!b]
\centering
\includegraphics[width=3in]{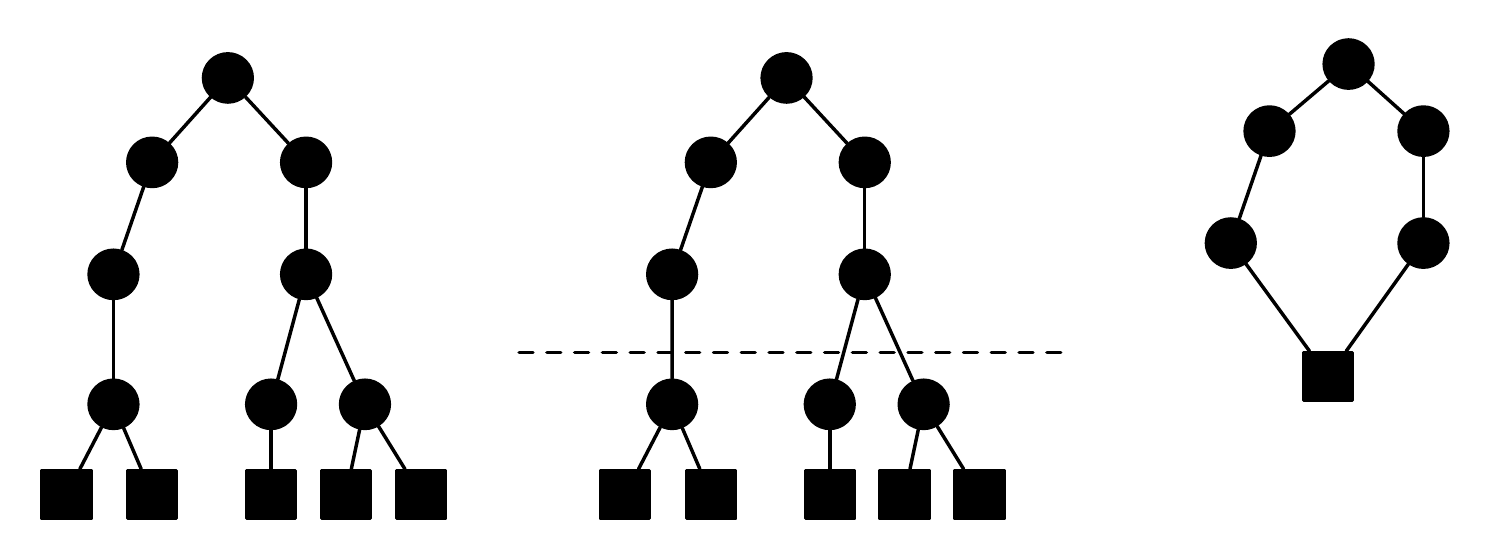}
\caption{Trie Truncation and Node Compression}
\label{fig:Truncation}
\end{figure}

In this manuscript, an additional step is added. The sparse matrix representing the trie is compressed using a Compressed Row Storage (CRS) algorithm, reducing furthermore the memory footprint of the bitmap array~\cite{wu2006optimizing}\cite{wang2014sparse}. The compressed row storage also allows the array to be stored in texture memory, hence benefiting from cached data. The row pointer generated by the CRS algorithm is stored in shared memory, benefiting from both the cache and an on-chip location reducing the clock cycles when accessing data. \newline

Figure~\ref{fig:Truncation} is a visual representation of steps I to V undertaken during the construction of the trie. As shown, the trie is truncated to an appropriate level. This technique was used by Vasiliadis~\textit{et~al.}~\cite{vasiliadis2010gravity} and further studied by Bellekens~\textit{et~al.}~\cite{bellekens2016high}. After truncation, similar suffixes within the trie are merged together and the leave nodes are replaced by a single end node. 
\begin{figure}[!b]
  \includegraphics[width=3in]{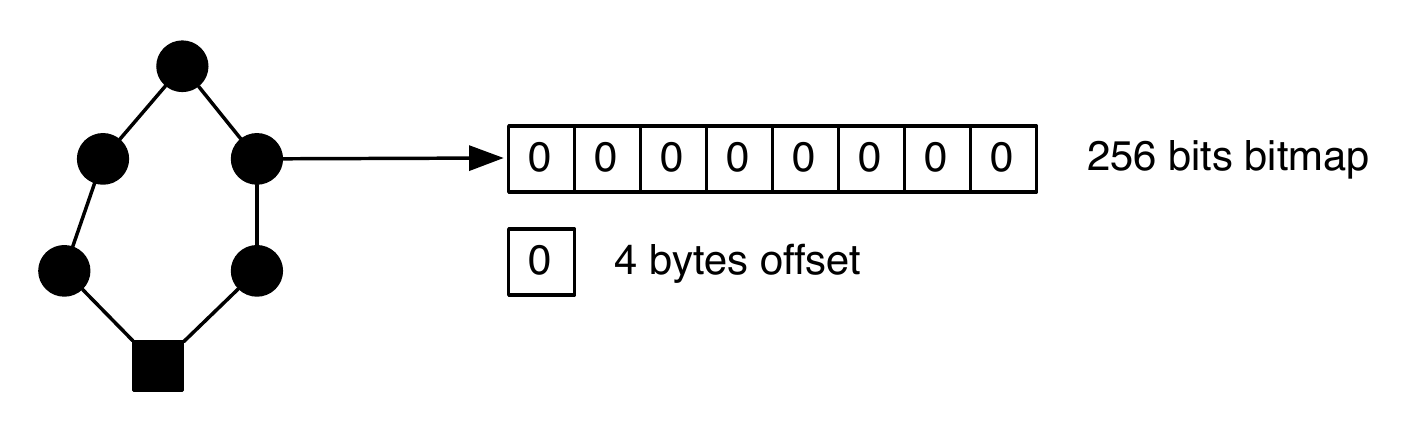}
  \caption{Bitmapped Nodes}
\label{fig:bitmapped}
\end{figure}
\newline Figure~\ref{fig:bitmapped} shows the composition of the nodes in the trie. Each node is composed of a bitmap of 256 bits from the ASCII alphabet. The bitmap is modular and can be modified based on the trie requirements (e.g., for DNA storage). Each node also contains an offset value providing the location of the first child of the current node. Subsequent children can be located following the method described in~\cite{bellekens2016high}. \newline

Figure~\ref{fig:prefixMatching} depicts four different memory layouts in order to achieve better compression and increase the throughput on GPUs.  Figure~\ref{fig:prefixMatching}~(A) represents the trie created with a linked list. Figure~\ref{fig:prefixMatching}~(B) represents the trie organised in a row major order, this allows the removal of pointers and simplifies the transition between the host and and the device memory. Figure~\ref{fig:prefixMatching}~(C) represents the trie in a two dimensional array, allowing the trie to be stored in Texture memory on the GPU and annihilate the trade-off between the compression highlighted in Figure~\ref{fig:prefixMatching}~A) and the throughput. Finally, Figure~\ref{fig:prefixMatching}~(D) is improving upon the compression of our prior research while allowing the trie to be stored in texture memory and the row\_ptr to be stored in shared memory. \newline

The CRS compression of the non-symmetric sparse matrix $A$ is achieved by creating three  vectors. The $val$ vector stores the values of the non-zero elements present within $A$, while the $col-ind$ store the indexes of the $val$. The storage savings for this approach is defined as $2nnz+n+1$, where $nnz$ represents the number of non-zero elements in the matrix and $n$ the number of elements per side of the matrix. In the example provided in Figure~\ref{fig:prefixMatching}~(C~and~D), the sparse matrix is reduced from 36 to 24 elements. \newline

The sparse matrix compression combined with the trie compression and the bitmap allows for storing large numbers of patters on GPUs allowing its use in big data applications, anti-virus and intrusion detection systems. Note that the CRS compression is pattern dependent, hence the compression will vary with the alphabet in use and the type of patterns being searched for.

\begin{figure}[!t]
  \includegraphics[width=3.5in,height=5in]{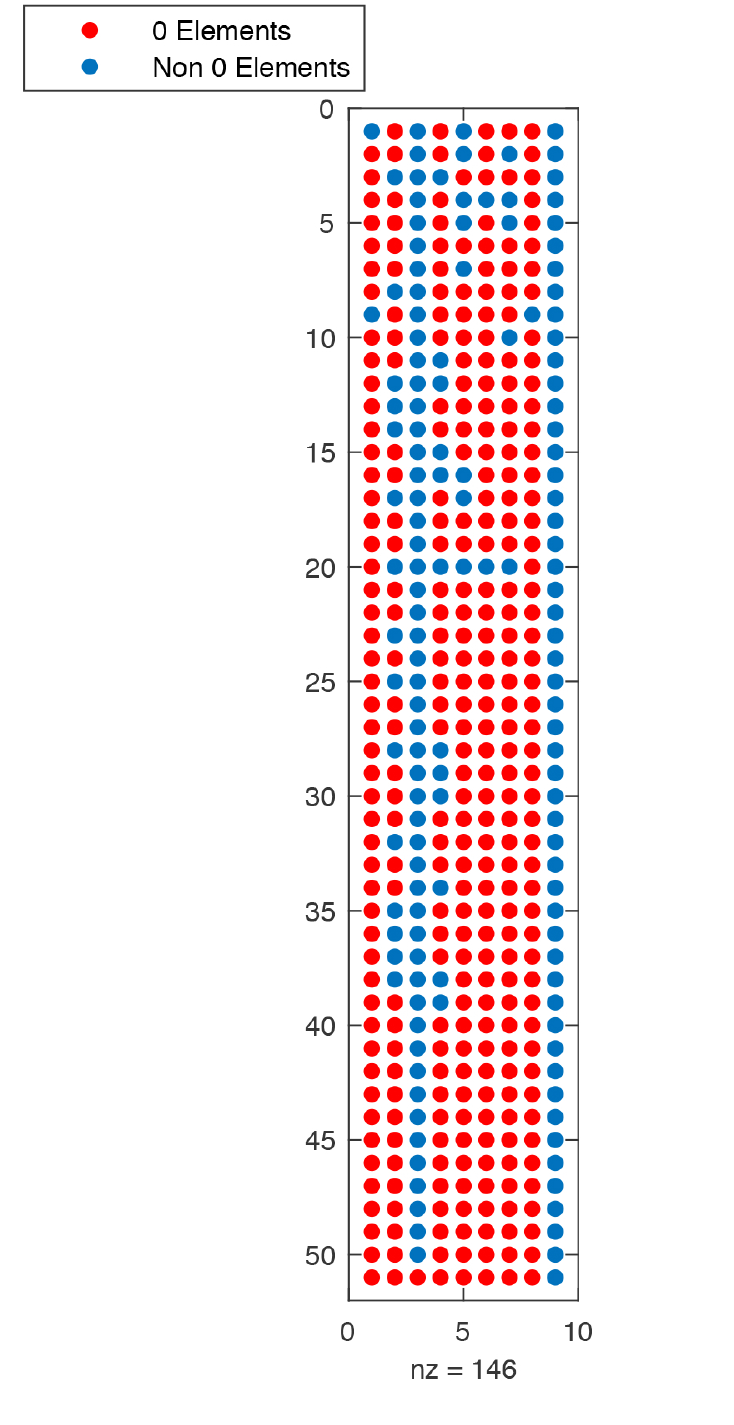}
  \caption{Sparse Matrix Representation of the a Compressed Trie Containing 10 Patterns}
\label{fig:SM}
\end{figure}
\section{Experimental Environment}
The Nvidia 1080 GPU is composed of 2560 CUDA cores divided into 20 SMs. The card also contains 8~GB~of~GDDR5X with a clock speed of 1733~MHz. Moreover the card possesses 96~KB shared memory and 48~KB of L1 cache, as well as 8 texture units and a GP104 PolyMorph engine used for vertex fetches for each streaming multiprocessors. The base system is composed of 2 Intel Xeon Processors with 20 cores, allowing up to 40 threads and has a total of 32GB of RAM. The server is running Ubuntu Server 14.04 with the latest version of CUDA 8.0 and C99.

\begin{figure}[!t]
  \includegraphics[width=3.5in]{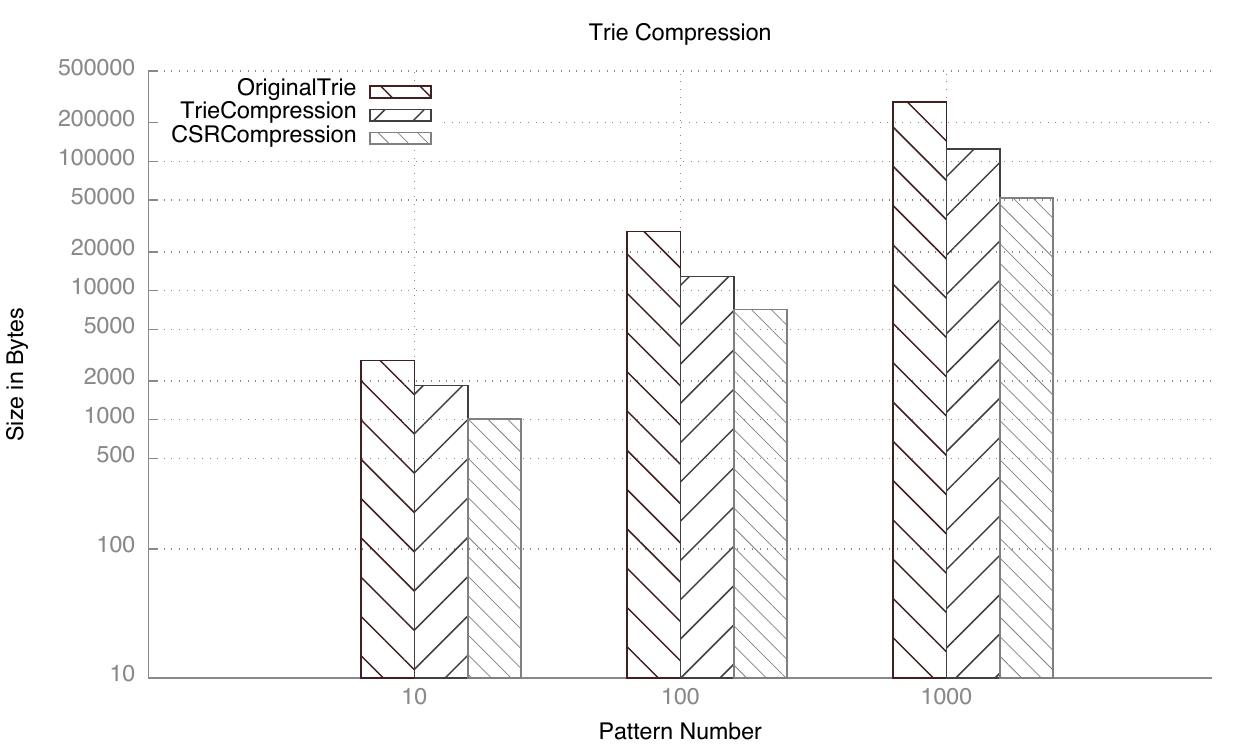}
  \caption{Comparison between the current state of the art and the CRS Trie}
\label{fig:CRS}
\end{figure}

\begin{figure}[!b]
  \includegraphics[width=3.5in]{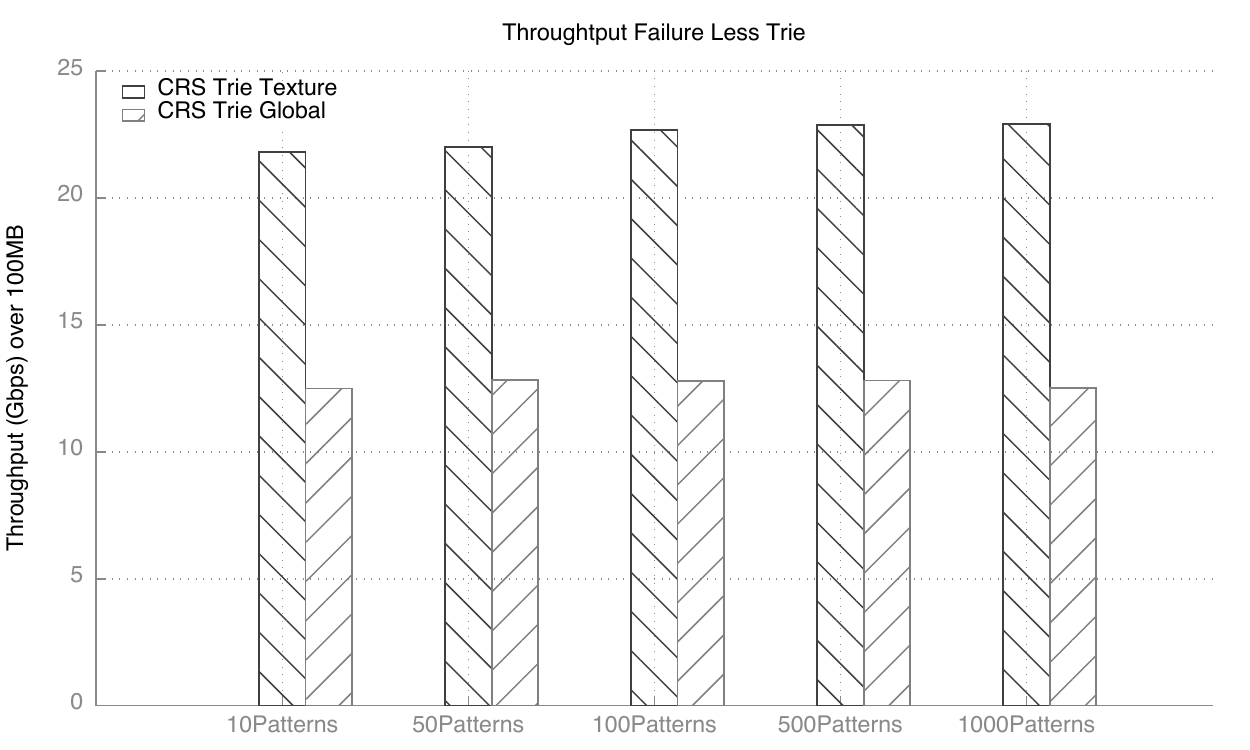}
  \caption{Throughput Comparison Between Global Memory and Texture Memory}
\label{fig:CRSThroughput}
\end{figure}

\section{Results}
The HEPFAC algorithm presented within this manuscript improves upon the state of the art   compression and uses texture memory and shared memory to increase the matching throughput.\newline

The evaluation of compression algorithm presented is made against the King James Bible. The patterns are chosen randomly within the text using the Mersenne Twister algorithm~\cite{Matsumoto:1998:MTE:272991.272995}. \newline

Figure~\ref{fig:SM} is a representation of the compressed trie stored in a 2D layout. The trie contains ten Patterns. The blue elements represent non-zero elements in the matrix while the red elements represent empty spaces within the matrix. The last column of the matrix only contains the offsets of each node. \newline

Figure~\ref{fig:CRS} demonstrates the compression achieved by the different compression steps aforementioned. The original trie required a total of 36 bytes for each nodes, 256 bits to represent the ASCII alphabet and four bytes for the offset. The trie compression, on the other hand requires 36 bytes for each node but reduces the size of the trie based on the alphabet used (in this case to eight levels), then merges similar suffixes together and merges all final nodes in a single one. Finally, the CRS compression algorithm compresses the sparse matrix representation of the trie. This technique allows an 83\% space reduction in comparison to the original trie and a 56\% reduction in comparison to the trie compression algorithm presented in~\cite{Bellekens:2014:HMS:2659651.2659723}. \newline

Figure~\ref{fig:CRSThroughput} depicts the throughput obtained when storing the CRS trie in global memory and in Texture memory. Global Memory does not provide access to a cache and requires up to 600 clock cycles to access data. This inherently limits the throughput of the pattern matching algorithm to 12~Gbps. When the CRS trie is stored in texture memory and the row\_ptr is stored in shared memory the algorithm demonstrate 22~Gbps throughput when matching a 1000 patterns within an alphabet $\Sigma=256$.

\section{Conclusion}
In this work a trie compression algorithm is presented. The trie compression scheme improves upon the state of the art and demonstrates 83\% space reduction against the original trie compression and 56\% reduction over the HEPFAC algorithm. Moreover, our approach also demonstrates over 22~Gbps throughput while matching a 1000 patterns. This work highlighted the algorithm on single GPU node, however, the algorithm can be adapted to cloud computing, or on FPGAs.

\bibliographystyle{IEEEtran}
\bibliography{bibliography}

\begin{thebibliography}{10}
\providecommand{\url}[1]{#1}
\csname url@samestyle\endcsname
\providecommand{\newblock}{\relax}
\providecommand{\bibinfo}[2]{#2}
\providecommand{\BIBentrySTDinterwordspacing}{\spaceskip=0pt\relax}
\providecommand{\BIBentryALTinterwordstretchfactor}{4}
\providecommand{\BIBentryALTinterwordspacing}{\spaceskip=\fontdimen2\font plus
\BIBentryALTinterwordstretchfactor\fontdimen3\font minus
  \fontdimen4\font\relax}
\providecommand{\BIBforeignlanguage}[2]{{%
\expandafter\ifx\csname l@#1\endcsname\relax
\typeout{** WARNING: IEEEtran.bst: No hyphenation pattern has been}%
\typeout{** loaded for the language `#1'. Using the pattern for}%
\typeout{** the default language instead.}%
\else
\language=\csname l@#1\endcsname
\fi
#2}}
\providecommand{\BIBdecl}{\relax}
\BIBdecl

\bibitem{nasridinov2014decision}
A.~Nasridinov, Y.~Lee, and Y.-H. Park, ``Decision tree construction on gpu:
  ubiquitous parallel computing approach,'' Computing, vol.~96, no.~5, 2014,
  pp. 403--413.

\bibitem{nakasato2009oct}
N.~Nakasato, ``Oct-tree method on gpu,'' arXiv preprint arXiv:0909.0541, 2009.

\bibitem{7576464}
S.~Memeti and S.~Pllana, ``Combinatorial optimization of work distribution on
  heterogeneous systems,'' in 2016 45th International Conference on Parallel
  Processing Workshops (ICPPW), Aug 2016, pp. 151--160.

\bibitem{aparicio2014scalable}
D.~Aparicio, P.~Paredes, and P.~Ribeiro, ``A scalable parallel approach for
  subgraph census computation,'' in European Conference on Parallel
  Processing.\hskip 1em plus 0.5em minus 0.4em\relax Springer International
  Publishing, 2014, pp. 194--205.

\bibitem{retvari2013compressing}
G.~R{\'e}tv{\'a}ri, J.~Tapolcai, A.~K{\H{o}}r{\"o}si, A.~Majd{\'a}n, and
  Z.~Heszberger, ``Compressing ip forwarding tables: towards entropy bounds and
  beyond,'' in ACM SIGCOMM Computer Communication Review, vol.~43, no.~4.\hskip
  1em plus 0.5em minus 0.4em\relax ACM, 2013, pp. 111--122.

\bibitem{bedorf2012bonsai}
J.~B{\'e}dorf, E.~Gaburov, and S.~P. Zwart, ``Bonsai: A gpu tree-code,'' arXiv
  preprint arXiv:1204.2280, 2012.

\bibitem{martinez2016practical}
M.~A. Mart{\'\i}nez-Prieto, N.~Brisaboa, R.~C{\'a}novas, F.~Claude, and
  G.~Navarro, ``Practical compressed string dictionaries,'' Information
  Systems, vol.~56, 2016, pp. 73--108.

\bibitem{navarro2016faster}
G.~Navarro and A.~O. Pereira, ``Faster compressed suffix trees for repetitive
  collections,'' Journal of Experimental Algorithmics (JEA), vol.~21, no.~1,
  2016, pp. 1--8.

\bibitem{tran2011bit}
T.~T. Tran, M.~Giraud, and J.-S. Varr{\'e}, ``Bit-parallel multiple pattern
  matching,'' in International Conference on Parallel Processing and Applied
  Mathematics.\hskip 1em plus 0.5em minus 0.4em\relax Springer, 2011, pp.
  292--301.

\bibitem{Aho:1975:ESM:360825.360855}
\BIBentryALTinterwordspacing
A.~V. Aho and M.~J. Corasick, ``Efficient string matching: An aid to
  bibliographic search,'' Commun. ACM, vol.~18, no.~6, Jun. 1975, pp. 333--340.
  [Online]. Available: \url{http://doi.acm.org/10.1145/360825.360855}
\BIBentrySTDinterwordspacing

\bibitem{vakili2016memory}
S.~Vakili, J.~Langlois, B.~Boughzala, and Y.~Savaria, ``Memory-efficient string
  matching for intrusion detection systems using a high-precision pattern
  grouping algorithm,'' in Proceedings of the 2016 symposium on architectures
  for networking and communications systems.\hskip 1em plus 0.5em minus
  0.4em\relax ACM, 2016, pp. 37--42.

\bibitem{bellekens}
X.~Bellekens, I.~Andonovic, R.~Atkinson, C.~Renfrew, and T.~Kirkham,
  ``Investigation of {GPU}-based pattern matching,'' in The 14th Annual Post
  Graduate Symposium on the Convergence of Telecommunications, Networking and
  Broadcasting (PGNet2013), 2013.

\bibitem{tumeo2010efficient}
A.~Tumeo, O.~Villa, and D.~Sciuto, ``Efficient pattern matching on gpus for
  intrusion detection systems,'' in Proceedings of the 7th ACM international
  conference on Computing frontiers.\hskip 1em plus 0.5em minus 0.4em\relax
  ACM, 2010, pp. 87--88.

\bibitem{5683320}
C.-H. Lin, S.-Y. Tsai, C.-H. Liu, S.-C. Chang, and J.-M. Shyu, ``Accelerating
  string matching using multi-threaded algorithm on {GPU},'' in Global
  Telecommunications Conference (GLOBECOM 2010), 2010 IEEE, Dec 2010, pp. 1--5.

\bibitem{Bellekens:2014:HMS:2659651.2659723}
\BIBentryALTinterwordspacing
X.~J.~A. Bellekens, C.~Tachtatzis, R.~C. Atkinson, C.~Renfrew, and T.~Kirkham,
  ``A highly-efficient memory-compression scheme for gpu-accelerated intrusion
  detection systems,'' in Proceedings of the 7th International Conference on
  Security of Information and Networks, ser. SIN '14.\hskip 1em plus 0.5em
  minus 0.4em\relax New York, NY, USA: ACM, 2014, pp. 302:302--302:309.
  [Online]. Available: \url{http://doi.acm.org/10.1145/2659651.2659723}
\BIBentrySTDinterwordspacing

\bibitem{bellekens2016high}
\BIBentryALTinterwordspacing
X.~J.~A. Bellekens, ``High performance pattern matching and data remanence on
  graphics processing units,'' Ph.D. dissertation, University of Strathclyde,
  2016. [Online]. Available:
  \url{http://ethos.bl.uk/OrderDetails.do?uin=uk.bl.ethos.698532}
\BIBentrySTDinterwordspacing

\bibitem{wu2006optimizing}
K.~Wu, E.~J. Otoo, and A.~Shoshani, ``Optimizing bitmap indices with efficient
  compression,'' ACM Transactions on Database Systems (TODS), vol.~31, no.~1,
  2006, pp. 1--38.

\bibitem{wang2014sparse}
H.~Wang, ``Sparse array representations and some selected array operations on
  gpus,'' Master's thesis, School of Computer Science University of the
  Witwatersrand A dissertation submitted to the Faculty of Science, University
  of the Witwatersrand, Johannesburg, 2014.

\bibitem{vasiliadis2010gravity}
G.~Vasiliadis and S.~Ioannidis, ``Gravity: a massively parallel antivirus
  engine,'' in International Workshop on Recent Advances in Intrusion
  Detection.\hskip 1em plus 0.5em minus 0.4em\relax Springer, 2010, pp. 79--96.

\bibitem{Matsumoto:1998:MTE:272991.272995}
\BIBentryALTinterwordspacing
M.~Matsumoto and T.~Nishimura, ``Mersenne twister: A 623-dimensionally
  equidistributed uniform pseudo-random number generator,'' ACM Trans. Model.
  Comput. Simul., vol.~8, no.~1, Jan. 1998, pp. 3--30. [Online]. Available:
  \url{http://doi.acm.org/10.1145/272991.272995}
\BIBentrySTDinterwordspacing

\end{thebibliography}

\end{document}